\documentclass[fleqn,10pt]{wlscirep}
\usepackage{braket}
\title{Discorrelated quantum states}

\author[1,*]{Evan Meyer-Scott}
\author[1]{Johannes Tiedau}
\author[1]{Georg Harder}
\author[2]{Lynden K. Shalm}
\author[1]{Tim J. Bartley}
\affil[1]{Department of Physics, University of Paderborn, Warburger Stra\ss e 100, 33098 Paderborn, Germany}
\affil[2]{National Institute of Standards and Technology, 325 Broadway, Boulder, CO, 80305, USA}

\affil[*]{evan.meyer.scott@upb.de}

\begin{abstract}
The statistical properties of photons are fundamental to investigating quantum mechanical phenomena using light. In multi-photon, two-mode systems, correlations may exist between outcomes of measurements made on each mode which exhibit useful properties. Correlation in this sense can be thought of as increasing the probability of a particular outcome of a measurement on one subsystem given a measurement on a correlated subsystem. Here, we show a statistical property we call ``discorrelation,'' in which the probability of a particular outcome of one subsystem is reduced to zero, given a measurement on a discorrelated subsystem. We show how such a state can be constructed using readily available building blocks of quantum optics, namely coherent states, single photons, beam splitters and projective measurement. We present a variety of discorrelated states, show that they are entangled, and study their sensitivity to loss. 
\end{abstract}
\begin{document}

\flushbottom
\maketitle

\thispagestyle{empty}

\section*{Introduction}

Quantum optics is, at its core, the study of the distributions of photons in modes of the electromagnetic field. These distributions can exhibit fundamental physical features such as photon antibunching~\cite{BROWN1956Correlat,PhysRevLett.39.691,RevModPhys.54.1061} and photon number squeezing~\cite{Loudon:1987aa,Kim:1989aa}, which cannot be explained using classical assumptions about the photon distribution. When we consider the joint distribution of photons across multiple modes, further nonclassical phenomena emerge, such as the presence of nonclassical correlations in the number of photons measured in each mode~\cite{PhysRevLett.59.2555,PhysRevLett.65.1551,PhysRevLett.69.2650,1742-6596-84-1-012020,PhysRevA.76.013833,PhysRevA.82.011801,Kalashnikov:12,Allevi:14,PhysRevA.91.043816,PhysRevLett.116.143601}. For an ideal two-mode squeezed vacuum state, the number of photons in each mode is always equal---the outcome of photon-number measurements are completely correlated~\cite{2014arXiv1401.4118L}. By contrast, when two indistinguishable photons are incident on different ports of a balanced beam splitter, bosonic bunching dictates that both photons will exit the same port~\cite{PhysRevLett.59.2044}, such that photon number measurements are anti-correlated. Here we introduce {\em discorrelation}, in which the joint probability $P_{n,n}$ of measuring $n$ photons in each mode is precisely zero for all $n$, but the marginal distributions $P_n$ are nonzero for all $n$. Discorrelation is distinct from correlation, anti-correlation, and decorrelation, extending our understanding of quantum correlations. It is an infinite-dimensional version of ``exclusive correlations'', here analysed as an effect of photon statistics rather than in the context of generalised Bell states~\cite{1367-2630-11-1-013006}. In this work we show ways to generate discorrelated states using commonly available input states and standard quantum optical techniques, and analyse the entanglement properties and loss behaviour of these states.

Correlation has been studied extensively in quantum optics in the context of communication, namely as a means to share common randomness between two parties~\cite{PhysRevLett.67.661,Bennett1992Quantum-,PhysRevLett.84.4729,PhysRevA.61.042302,PhysRevA.63.052311,Devetak:2004aa,PhysRevLett.102.130501}. In this context, discorrelation can be used to share {\em unique} randomness between parties, complementary to conventional quantum communication protocols. Unique randomness, where each party has a random number that is distinct from the other parties', could be useful in distributed voting schemes~\cite{Bohli2007} or for fairly dealing cards~\cite{Golle:2005aa} in the area of cryptographic study known as ``mental poker''~\cite{Shamir1981}. Mental poker and distributed voting involve the allotment of cards, voter identifiers, or other pieces of information fairly and secretly. By ensuring each individual receives a unique random number without knowledge of any others, discorrelation could remove the need for a trusted third party, replacing it with the fundamental randomness of quantum superpositions.

We propose two methods for generating discorrelation. The first is the displacement of a single photon by a coherent state on a beam splitter, producing an entangled state~\cite{PhysRevA.65.032323,PhysRevA.66.024303,Lvovsky:2013aa,Bruno2013Displace}. In this case, the discorrelated state is generated locally and then shared between parties. Although various aspects of these and related states have been analysed~\cite{Windhager:2011ab,PhysRevA.86.060301,PhysRevA.93.023845}, for example in the context of micro-macro entanglement~\cite{PhysRevA.85.032307} and N00N-state generation~\cite{PhysRevA.76.031806,PhysRevLett.100.073601}, the joint photon number distribution and the discorrelation therein, has not.

\begin{figure}[h!pt]
\centerline{\includegraphics[width=0.5\columnwidth]{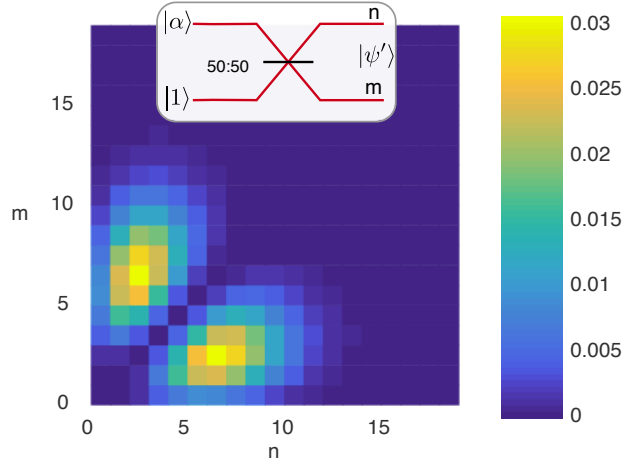}}
\caption{Interfering a single photon with a coherent state ($\alpha = \sqrt{8}$) on a 50:50 beam splitter (inset) produces discorrelation: the photon number at the two output ports $m$ and $n$ (shown as a heatmap) can take any value individually, but together the two ports can never produce $m = n$.}
\label{fig:coh_oneph_dist}
\end{figure}

In the second case, we show how discorrelated multidimensional photon statistics can be generated nonlocally using a single shared two-dimensional state. This method is based on the coherent superposition of photon addition and subtraction, which has been proposed for generating nonclassical states~\cite{PhysRevLett.88.250401,Browne:2004aa,PhysRevLett.98.203602,PhysRevA.82.053812,PhysRevA.84.012302,PhysRevA.86.043820} and distilling entanglement in continuous-variable quantum states~\cite{PhysRevA.87.032307,1367-2630-17-2-023038}. As with related photon subtraction and addition experiments~\cite{Zavatta660,PhysRevA.73.042310,Parigi1890,PhysRevLett.98.030502,0953-4075-41-13-133001,Yang:09,Takahashi:2010aa,PhysRevA.87.022313}, and in photon-number-difference filtering~\cite{PhysRevA.86.063823}, photon detections are used to perform nonlinear operations, but in our case the input states can begin separable and become entangled by the discorrelation operation.  The two methods for creating discorrelation are closely related as they both rely on the modification of photon number distributions due to the interference of Fock states with other continuous-variable states.

\section*{Results}

The simplest way to create a discorrelated state (Fig.~\ref{fig:coh_oneph_dist}) is to impinge a single photon $\ket{1}$ and a coherent state $\ket{\alpha}$ on two ports of a 50:50 beam splitter. The coherent state displaces the single photon~\cite{PhysRevA.86.060301}, resulting in the entangled two-mode state 
\begin{align}\label{eqn:pnbasis}
\ket{\psi^\prime}=\mathcal{N}\sum_{n=0,m=0}^\infty c^\prime_{n,m}\ket{n,m},
\end{align} 
where the coefficients are given by 
\begin{equation}
c^\prime_{n,m} = \frac{\left(\tfrac{\alpha}{\sqrt{2}}\right)^{n+m-1}}{\sqrt{2n!m!}} (n-m),
\end{equation}
and the normalization factor is $\mathcal{N}=e^{-\frac{\left|\alpha\right|^2}{2}}$.
Thus the two output modes will have the same marginal photon number distribution, but since photon number measurements on the two modes will never give the result $n = m$, the output state is discorrelated. We show this discorrelated photon number distribution in Fig.~\ref{fig:coh_oneph_dist} as a heatmap of the joint photon number detection probabilities.

\subsection*{Nonlocal, adaptable discorrelation}

\begin{figure}[thp]
\centerline{\includegraphics[width=.5 \columnwidth]{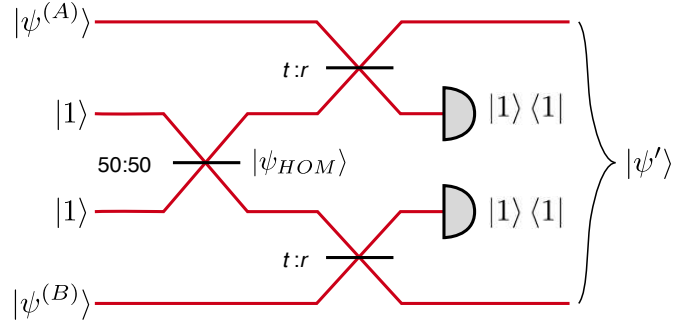}}
\caption{Schematic for generating discorrelated states nonlocally. Two independent multiphoton states $\ket{\psi^{(A)}}$ and $\ket{\psi^{(B)}}$ each interfere with one mode of the entangled state $\ket{\psi_{HOM}}$, generated by HOM interference of single photons. The resulting two-mode state $\ket{\psi^\prime}$ may become discorrelated dependent on the phase relationship between the two input stats.}
\label{fig:moses_setup}
\end{figure}

One can also implement the discorrelation operation in a nonlocal manner, whereby particular projective measurements can be used to herald states whose photon distributions depend on the parameters of the interaction. We employ an entangled HOM state~\cite{PhysRevLett.59.2044}, created by the interference of two single photons at a 50:50 beam splitter, to distribute entanglement between two parties. As seen in Fig.~\ref{fig:moses_setup}, the HOM state $\ket{\psi_{HOM}}=\tfrac{1}{\sqrt{2}}\left(\ket{20}+\ket{02}\right)$ interferes with two other (separable) multiphoton states, expressed in the photon number basis as $\ket{\psi^{(A)}}=\sum_{n=0}^\infty c^{(A)}_n\ket{n}$ and $\ket{\psi^{(B)}}=\sum_{m=0}^\infty c^{(B)}_m\ket{m}$, where the coefficients $c^{(A)}_n,~c^{(B)}_m$ are normalized and the states share a phase reference. Following projective measurements of single photons at one output of each beam splitter (of transmissivity $t$, reflectivity $r$), the entanglement in the HOM state is mapped to the (pure) output state $\ket{\psi^\prime}$. This allows the two parties to share a state whose discorrelation depends on the input states and beam splitter parameters, without interacting directly. The coefficients of the output state equation~(\ref{eqn:pnbasis}) are now given by
\begin{align}
c_{n,m}^\prime=&\left(-1\right)^{m}t^2r^{n+m-2}\label{eqn:general_state}
\times\Big\{c^{(A)}_{n+1} c^{(B)}_{m-1}\sqrt{m\left(n+1\right)}\left[1-\frac{\left(m+1\right)t^2}{2}\right]-c^{(A)}_{n-1} c^{(B)}_{m+1}\sqrt{n\left(m+1\right)}\left[1-\frac{\left(n+1\right)t^2}{2}\right]\Big\}.
\end{align}
Now the normalization factor $\mathcal{N}$ is related to the probability of the measurement of the two ancilla photons $P_H$, calculated by
\begin{equation}
P_H=\mathcal{N}^{-2}=\sum_{n,m=0}^\infty |c^\prime_{n,m}|^2.
\end{equation}

Given the projective measurement of single photons by the two ancilla detectors, the probability $P_{n,m}$ of a particular joint measurement of photon number $n,m$ on the remaining modes is given by $P_{n,m}=\mathcal{N}^{2}|c^\prime_{n,m}|^2$. The condition for discorrelation between the two parties is that $P_{n,n}=0$ for all $n$, therefore we seek solutions to $|c^\prime_{n,n}|^2=0$ which are independent of $n$. By setting $m=n$ in equation~(\ref{eqn:general_state}) we obtain
\begin{align}\label{eqn:general_state_mn}
c^\prime_{n,n}= \left(-1\right)^{n}t^2r^{2\left(n-1\right)}\sqrt{n\left(n+1\right)}
\times\left[1-\frac{\left(n+1\right)t^2}{2}\right]\Big(c^{(A)}_{n+1} c^{(B)}_{n-1}-c^{(A)}_{n-1} c^{(B)}_{n+1}\Big).
\end{align}

This can be set to zero by setting either term in parentheses to zero. The term $\left[1-\left(n+1\right)\tfrac{t^2}{2}\right]$ can be set to zero if $t=\sqrt{\tfrac{2}{n+1}}$. However, this does not satisfy our condition for discorrelation due to the dependence on $n$. Instead, it is the two-mode analogue of filtering out photonic Fock states for entanglement generation, which has been demonstrated for a single mode~\cite{PhysRevLett.98.203602}. 

The second term in parentheses in equation~(\ref{eqn:general_state_mn}) is zero if
\begin{equation}
c^{(A)}_{n+1}c^{(B)}_{n-1}=c^{(A)}_{n-1} c^{(B)}_{n+1}~,
\end{equation}
or alternatively
\begin{equation}\label{eqn:discorrelation_condition}
\frac{c^{(A)}_{n+1}}{c^{(A)}_{n-1}}=\frac{c^{(B)}_{n+1}}{c^{(B)}_{n-1}}.
\end{equation}
\begin{figure}[thp]
\centerline{\includegraphics[width= \columnwidth]{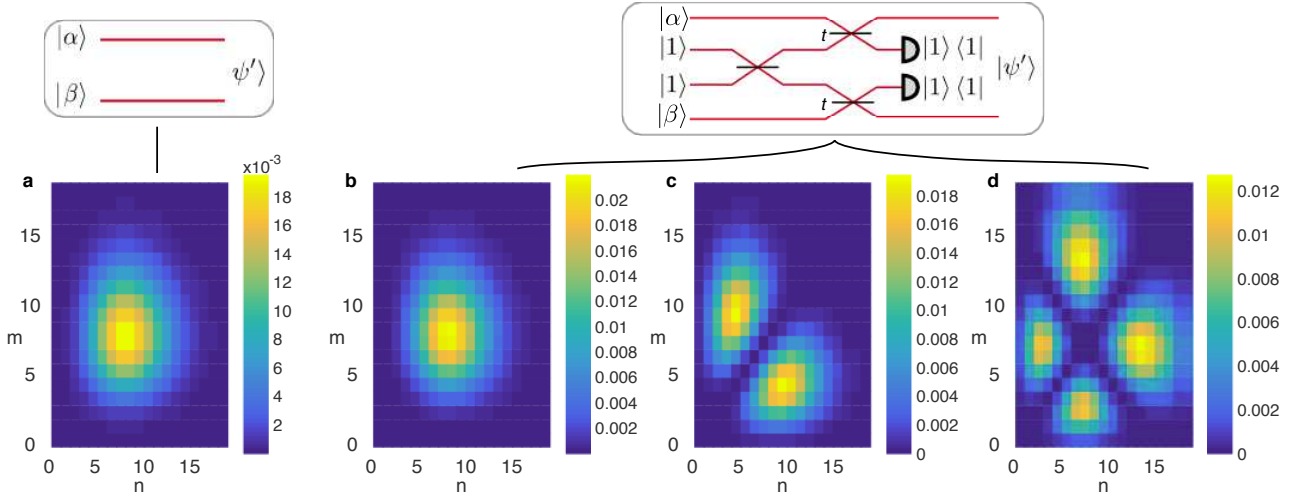}}
\caption{Probability distributions of photon numbers $(n,m)$ (a) of the input modes before the discorrelation procedure and (b-d) of the output modes ($\ket{\psi^\prime}$ in Fig.~\ref{fig:moses_setup}). Each case corresponds to different relationship between the phases of the coherent states and the beam splitter transmissivity: (b) $\alpha=\sqrt{8}=-i\beta,~t=\sqrt{\tfrac{2}{30}}$, (c) $\alpha=\beta,~t=\sqrt{\tfrac{2}{30}}$, (d) $\alpha = \beta,~t=\sqrt{\tfrac{2}{15}}$. In the latter two cases the terms with $n = m$ are eliminated, and in the third case, terms with $n+m = 14$ are also removed. The logarithmic negativities of the four states (a measure of entanglement) are 0, 0.04, 1, and 1 respectively.}
\label{fig:Pnmall}
\end{figure}
If this condition is met, there is precisely zero probability that the same photon number is measured on each of the two modes, corresponding to a completely discorrelated state. This condition is independent of the beam splitter transmissivity $t$, although we assume it is the same in both modes. Whether the output state is discorrelated therefore depends on the two initial states $\ket{\psi^{(A)}}$ and $\ket{\psi^{(B)}}$. In fact, changing only one of the input states allows to turn on or off the discorrelation, as seen in the examples below.

\subsection*{Discorrelation with coherent states}

The input coefficients for coherent states $\ket{\psi^{(A)}}=\ket{\alpha}$ and $\ket{\psi^{(B)}}=\ket{\beta}$ are
\begin{align}
c_n^{(A)}=e^{-\frac{|\alpha|^2}{2}}\frac{\alpha^n}{\sqrt{n!}},~~
c_{n^\prime}^{(B)}=e^{-\frac{|\beta|^2}{2}}\frac{\beta^{n^\prime}}{\sqrt{n^{\prime}!}},
\end{align}
where $\alpha=|\alpha|e^{i\phi_A}$ and $\beta=|\beta|e^{i\phi_B}$ are the complex amplitudes of each coherent state. To fulfill the condition in equation~(\ref{eqn:discorrelation_condition}), we find that $\alpha^2=\beta^2$, i.e. that the two coherent states must have the same amplitude with integer multiples of $\pi$ phase between them. The coefficient $c_{n,m}^\prime$ in equation~(\ref{eqn:general_state}) thus reduces to
\begin{align}
c_{n,m}^\prime=e^{-|\alpha|^2}\frac{\alpha^{n+m}}{\sqrt{n!m!}}\left(-1\right)^mt^2r^{n+m-2}
\left[1-\frac{t^2}{2}\left(n+m+1\right)\right]\left(n-m\right),\label{eqn:cohdiscorr}
\end{align}
from which it is clear from the last term in parentheses that $c_{n,m}^\prime=0$ when $n=m$, independent of all other parameters including the magnitude of the coherent state $|\alpha|$.

We show in Fig.~\ref{fig:Pnmall} the joint photon number probabilities corresponding to this interaction, which we calculated using the Quantum Optics Toolbox~\cite{1464-4266-1-4-312}. Depending on the relationship between the phases of the coherent states and the transmissivity of the beam splitter, a wide variety of exotic number distributions can be generated, and in particular discorrelation can be turned on or off by varying the relative phase between $\alpha$ and $\beta$. Tuning the beam splitter transmission also allows eliminating terms with $n+m = \tfrac{2}{t^2}-1$ due to the term in square brackets in Eq~(\ref{eqn:cohdiscorr}). Even with no discorrelation as in Fig.~\ref{fig:Pnmall}(b), the state is still slightly entangled by the operation, despite the number distribution being extremely similar to the unentangled input states of Fig.~\ref{fig:Pnmall}(a). Unlike the discorrelated states in (c) and (d), however, the amount of entanglement as quantified by the logarithmic negativity~\cite{PhysRevLett.95.090503} depends on the coherent state strength $\alpha$ and beam splitter transmissivity $t$.

\subsection*{Discorrelation with squeezed vacuum}

\begin{figure*}[htp]
\centerline{\includegraphics[width= \columnwidth]{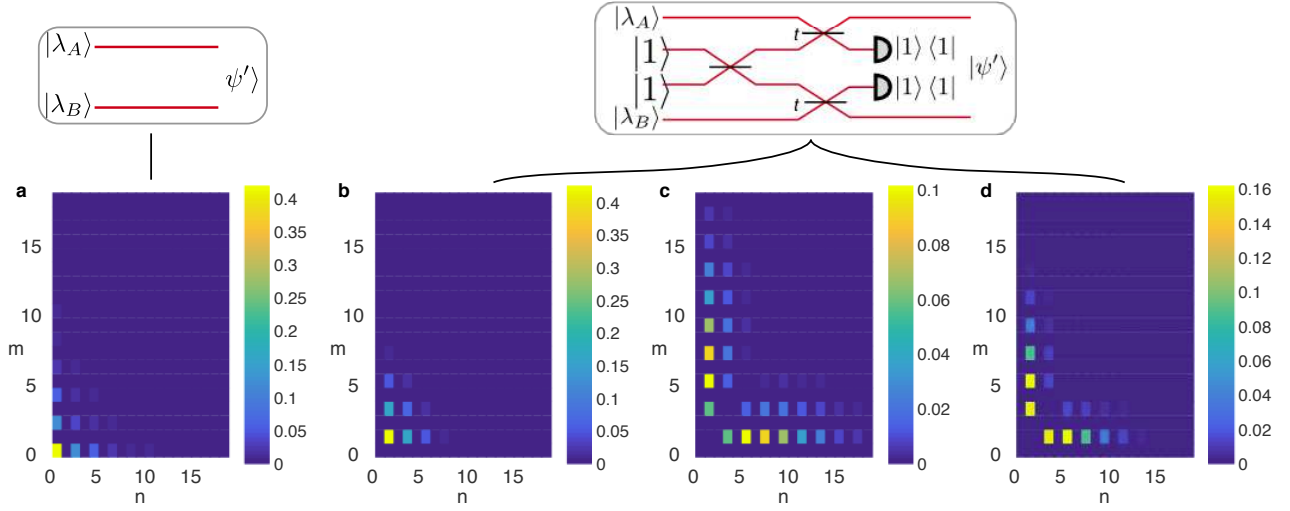}}
\caption{Probability distribution of different combinations of photon numbers $(n,m)$ (a) of the input modes before the discorrelation procedure and (b-d) of the output modes with single-mode squeezed vacuum inputs. Each case corresponds to different relationship between the phases of the SMSV and the beam splitter transmissivity: (b) $\lambda_1=1=-\lambda_2,~t=\sqrt{\tfrac{2}{9}}$, (c) $\lambda_1=\lambda_2,~t=\sqrt{\tfrac{2}{9}}$, (d) $\lambda_1=\lambda_2,~t=\sqrt{\tfrac{2}{5}}$. In the latter two cases the terms with $n = m$ are eliminated, but here the components near $n+m = \tfrac{2}{t^2}$ tend to be amplified, rather than suppressed as for the coherent state input.}
\label{fig:SMSV}
\end{figure*}
We next consider two single-mode squeezed vacuum states $\ket{\lambda_{A}}$ and $\ket{\lambda_B}$, written in the number basis as 
\begin{align}
\ket{\lambda_A}=\mathcal{N}\sum_{n=0}^\infty\lambda_{A}^n\sqrt{\frac{\left(2n\right)!}{2^nn!}}\ket{2n},~~
\ket{\lambda_B}=\mathcal{N}\sum_{n=0}^\infty\lambda_{B}^n\sqrt{\frac{\left(2n\right)!}{2^nn!}}\ket{2n}.
\end{align}
 Here the   $c^{(A,B)}_{2n}=\lambda_{A,B}^n\sqrt{\frac{\left(2n\right)!}{2^nn!}}$ are unnormalized, with the squeezing parameters $0\leq|\lambda_{A,B}|<1$. Substituting these in the discorrelation criterion equation~(\ref{eqn:discorrelation_condition}) we find $\lambda_A = \lambda_B$; therefore if the two single-mode squeezed states have squeezing parameters with the same magnitude and phase, the resulting state will be discorrelated. In this case, the photon number distributions are shown in Fig.~\ref{fig:SMSV}, and just as with the coherent states, terms with $m=n$ are eliminated.

We can also apply the discorrelation operation to an entangled state. Now the input state is no longer separable, and can be written as
\begin{equation}
\ket{\psi_\textrm{in}}=\sum_{n,m=0}^\infty c_{n,m} \ket{n,m}.
\end{equation}
With the same discorrelation procedure as before, we generate the state
\begin{align}\nonumber
\ket{\psi_\textrm{out}}&=\mathcal{N}\sum_{n,m=0}^\infty t^2r^{n+m}\left(-1\right)^m\bigg\{c_{n-1,m+1}\sqrt{n\left(m+1\right)}\left[1-\frac{t^2}{2r^2}\left(n-1\right)\right]-c_{n+1,m-1}\sqrt{m\left(n+1\right)}\left[1-\frac{t^2}{2r^2}\left(m-1\right)\right]\bigg\}\ket{n,m}.\nonumber
\end{align} 
Again, this state is discorrelated if there exisits zero probability of measuring $\ket{n,n}\!\bra{n,n}$, which is the case when
\begin{equation}\label{eqn:ent_discorrelation_condition}
c_{n-1,n+1}=c_{n+1,n-1}.
\end{equation}

\begin{figure*}[htp]
\centerline{\includegraphics[width=.75 \columnwidth]{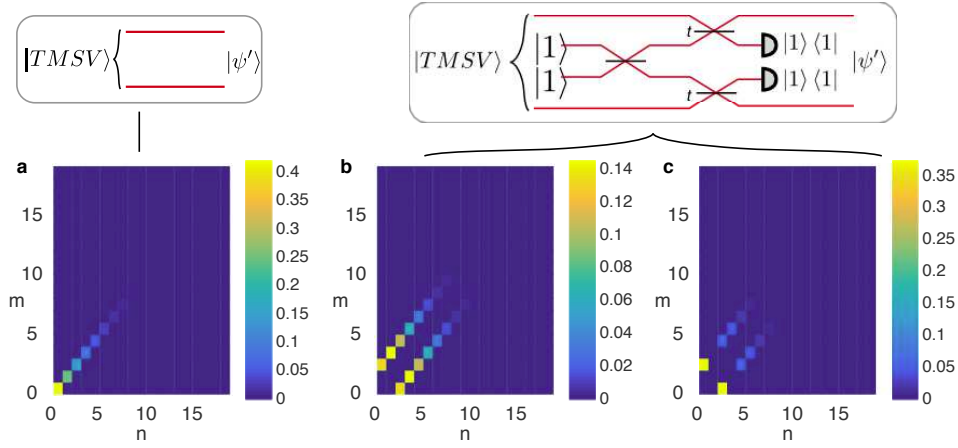}}
\caption{Probability distribution of different combinations of photon numbers $(n,m)$ (a) at the input modes and (b-c) output modes for a two-mode squeezed vacuum input with $\lambda = 1$. The latter two cases correspond to different beam splitter transmissivity: (b) $t=\sqrt{\tfrac{2}{15}}$, (c) $t=\tfrac{1}{\sqrt{2}}$. In both cases terms with $m=n$ are shifted to $n = m \pm 2$, and the latter also removes $n+m = 4$.}
\label{fig:TMSV}
\end{figure*}

We illustrate this with the case of a two-mode squeezed vacuum (TMSV) state. TMSV states exhibit perfect photon number correlations, which we modify with the discorrelation operation. Writing the TMSV state in the number basis as $\ket{TMSV}=\mathcal{N}\sum_{n=0}^\infty\lambda^n\ket{n,n}$ gives the coefficients $c_{n,m}=\lambda^n\delta_{n,m}$. The criterion equation~(\ref{eqn:ent_discorrelation_condition}) is fulfilled, since $c_{n-1,n+1}=c_{n+1,n-1}=0$, independent of the squeezing parameter $\lambda$ and any phases involved. 
The joint photon number distribution for this case is shown in Fig.~\ref{fig:TMSV}. Now the outputs are still tightly correlated, but with photon numbers offset by 2. For example, if $n = 2$, $m = 0$ or $m = 4$, as opposed to the input TMSV state where $n =m$.

\subsection*{Loss dependence}
\begin{figure}[htp]
\centerline{\includegraphics[width= \columnwidth]{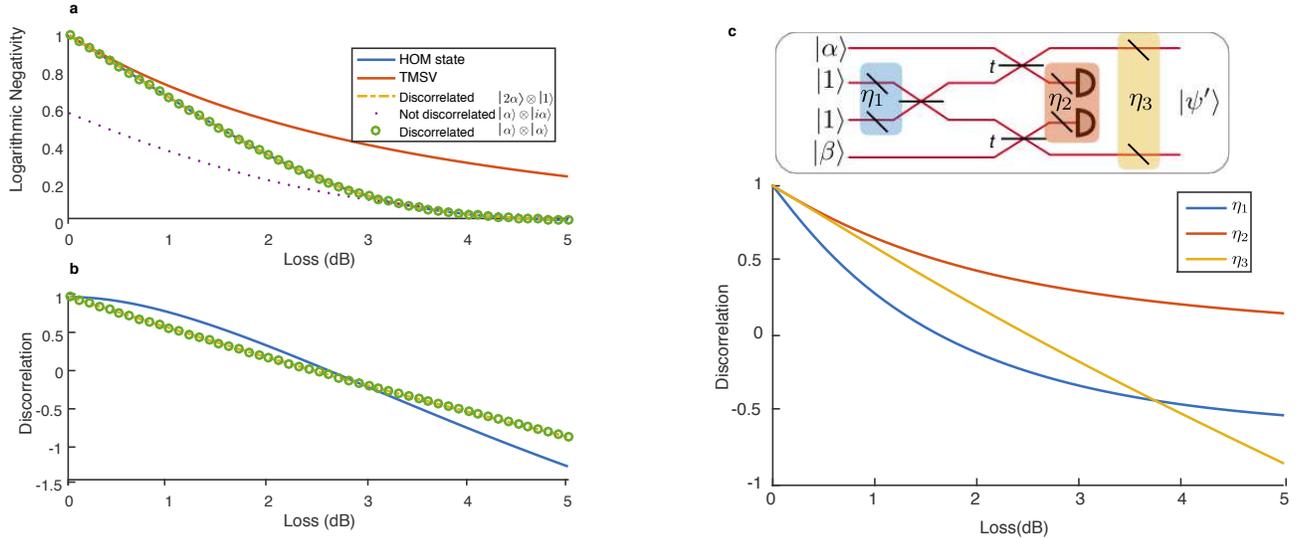}}
\caption{(a) Logarithmic negativity versus loss for  two-mode discorrelated states and other entangled states. The HOM state, a discorrelated state based on a displaced single photon (Fig.~\ref{fig:coh_oneph_dist}), and a discorrelated state based on displacing the HOM state (Fig.~\ref{fig:Pnmall}(c)) all show the same loss scaling, with the TMSV slightly better. The not-discorrelated state (Fig.~\ref{fig:Pnmall}(b), here with $t = 0.5$) has less entanglement, as it is closer to the unentangled input states. (b) Discorrelation versus loss for the three discorrelated states. A discorrelation $D<1$  indicates a nonzero probability of measuring the same outcome, and $D<0$ means the outcomes are more likely to be correlated than uncorrelated coherent states. The discorrelation for the TMSV is not plotted because it is always more correlated than the reference state, and thus has a large negative discorrelation ($\approx-3.5$). (c) Discorrelation versus loss at three different points in the circuit. In each case the loss is applied symmetrically to both arms. Discorrelation is most sensitive to loss in the ancilla photons before interference, and least sensitive to loss in the heralding detectors.}
\label{fig:logneg_loss}
\end{figure}

As in all protocols involving photon number statistics, it is important to study the effects of loss in realistic implementations~\cite{Wang:11}.  We show in Fig.~\ref{fig:logneg_loss}(a) the entanglement of three types of discorrelated states via the logarithmic negativity~\cite{PhysRevLett.95.090503} of the joint output state as a function of loss applied symmetrically to both modes of the state. All three states start with a logarithmic negativity of 1, and decay identically with loss. For comparison we show the two-mode squeezed vacuum with the squeezing parameter for a logarithmic negativity of 1 with no loss, and the ``not-discorrelated'' state from Fig.~\ref{fig:Pnmall}(b) (but with $t=0.5$), whose logarithmic negativity is not always 1 in the lossless case, but instead depends on the strength of the coherent states and the beam splitter transmissivity. For example, this not-discorrelated state has logarithmic negativity $\approx 1$ when $t=2/\alpha^2$, and $\approx 0$ when $t\ll 2/\alpha^2$.

As another quantification of the effect of loss, we introduce the discorrelation
 \begin{equation}
D = 1 - \frac{P_{m=n,discorr}}{P_{m=n,uncorr}},
\end{equation}
defined by the probability of observing the same photon number between the two parties compared to this probability for uncorrelated coherent states. Thus when $P_{m=n,discorr}=0$, $D = 1$ and when $P_{m=n,discorr}=P_{m=n,uncorr}$, $D = 0$. Both probabilities are calculated from normalized states, with the uncorrelated state additionally lossless. A negative discorrelation means the output state is more likely than uncorrelated states to produce a correlated result. For high loss, the ``correlated result'' is just the vacuum. We show in Fig.~\ref{fig:logneg_loss}(b) the loss behavior of discorrelation for three initally discorrelated states. We also analyse the effect of loss at various points in the discorrelation circuit: in the ancilla preparation, before the single-photon detectors, and after discorrelation. In the lossless case the state is perfectly discorrelated as seen in Fig.~\ref{fig:logneg_loss}(c), which falls off as loss is added.

\section*{Discussion}

We have presented a form of quantum correlation, {\em discorrelation}, with the property that joint measurements of the photon number by two parties never yield the same result. Discorrelation can be produced by interfering a single photon with a coherent state, or by interfering two coherent states with a Hong-Ou-Mandel entangled state plus photon detection, the latter of which allows tuning additional properties of the state. 

Discorrelation is a multideminsional phenomenon that maps the entanglement of the two-photon HOM state to much larger states. In fact the statistics of discorrelation can be interpreted as a generalised HOM-type bosonic bunching effect for higher numbers of photons, or as a displacement of the HOM state $\tfrac{1}{\sqrt{2}}\left(\ket{2,0}-\ket{0,2}\right)$, which retains a similar quantum signature in the photon-number basis. 
Our procedure is bears a practical resemblance to generating discrete-continuous hybrid entanglement~\cite{Jeong:2014aa,Morin:2014aa,PhysRevLett.114.100501}, although here we map, rather than swap, discrete-variable path entanglement to continuous-variable entanglement.

It may be possible to extend discorrelation to more than two modes, wherein each mode has a large distribution over photon number, but no two modes can have the same measurement result. This form of discorrelation could be used to share unique random numbers across many parties, which may be useful in an untrusted card dealer scenario. However, the security implications need further analysis, along with considerations on the scaling of such a protocol.

Our results indicate that discorrelated states can be generated from a variety of input states and with a variety of output statistics. However these discorrelated states, similar to HOM states, are not robust against photon loss. In addition, it will be important in experiments to consider the quality of the single photon ancillae and indistinguishability of the four modes. 


\section*{Acknowledgements}
We acknowledge support from the Initial Training Network PICQUE under the FP7 Marie Curie Actions of the European Commission (Grant No. 608062), the Natural Sciences and Engineering Research Council of Canada, and the Deutsche Forschungsgemeinschaft (DFG) via SFB/TRR142.

\section*{Author contributions statement}
T.B. conceived the idea and supervised the research. T.B., E.M.S., and L.K.S. developed the idea. E.M.S. performed numerical simulations with support from J.T. and G.H. The manuscript was written by E.M.S. and T.B with input from all authors. 

\section*{Additional information}
\textbf{Competing financial interests:}
The authors declare no competing financial interests.

\end{document}